\documentclass[reprint,floats,floatfix,amsmath,amssymb,nofootinbib,longbibliography]{revtex4-2}
\usepackage{amsfonts}
\usepackage{graphicx}
\usepackage{dcolumn}
\usepackage{bm}
\usepackage{natbib}
\usepackage{wasysym}
\usepackage[hyperindex,colorlinks]{hyperref}

\newcommand{\p}{\partial}
\newcommand{\rs}{\rho^{*}}
\newcommand{\bs}{\boldsymbol}
\newcommand{\bx}{\bar{x}}

\renewcommand{\H}{\hat{\mathcal{H}}}
\renewcommand{\a}{\hat{\alpha}}
\newcommand{\G}{\mathcal{G}^{(4)}}
\newcommand{\PhiG}{\Phi_{\mathcal{G}}}
\newcommand{\s}{\scriptscriptstyle}

\newcommand{\dal}{\Box \Phi}
\newcommand{\dpp}{\left(\nabla \Phi \right)^2}

\begin{document}
\title{A Post-Newtonian Analysis of Regularized 4D-EGB Theory:
Complete Set of PPN Parameters and Observational Constraints}

\author{Júnior D. Toniato}
 \email{junior.toniato@ufes.br}
\affiliation{Departamento de Química e Física - Centro de Ciências Exatas, Naturais e da Saúde, Universidade Federal do Espírito Santo - Campus Alegre, ES, 29500-000, Brazil.}%
\affiliation{Núcleo de Astrofísica e Cosmologia - Cosmo-Ufes, Universidade Federal do Espírito Santo, Vitória, ES,  29075-910, Brazil.}

\author{Mart\'in G. Richarte}
\email{martin@df.uba.ar}
\affiliation{Departamento de F\'isica - Universidade Federal do Esp\'irito Santo, 29075-910 Vit\'oria, ES, Brazil}
\affiliation{PPGCosmo, CCE - Universidade Federal do Esp\'irito Santo, 29075-910 Vit\'oria, ES, Brazil}
\affiliation{Departamento de F\'isica, Facultad de Ciencias Exactas y Naturales,
Universidad de Buenos Aires, Ciudad Universitaria 1428, Pabell\'on I, Buenos Aires, Argentina}

\date{\today}

\begin{abstract}
We performed a post-Newtonian analysis of the regularized four-dimensional Einstein-Gauss-Bonnet gravitational theory (4D-EGB). The resulting metric differs from the classical parametrized post-Newtonian (PPN) formalism in that a new gravitational potential arises from the integration of the approximate field equations. We also investigated the conserved quantities and equations of motion for massive bodies and light rays to a certain degree. By computing the predicted periastron advance rate in a binary system, we obtained an observational constraint that is stronger than those of previous analyses. Although the usual 10 PPN parameters can still be derived within the PPN framework, an extra parameter is needed to account for the full post-Newtonian tests.
\end{abstract}

\maketitle

\section{Introduction}
Einstein's theory of general relativity (GR) has passed numerous experimental tests since its inception in the early 20th century. GR could accurately explain the precession of Mercury's perihelia, a significant accomplishment at the time. Over the past century, GR has been put to the test on various fronts \cite{Will:2014kxa, Ishak:2018his}. In 2015, a major milestone was achieved when the LIGO interferometer successfully detected the first signals of gravitational waves (GWs)  emitted by the merger of two black holes \cite{LIGOScientific:2016aoc}. This groundbreaking discovery was followed by another significant event in 2017, when Advanced LIGO and Virgo detected the electromagnetic counterpart of a binary neutron star coalescence \cite{LIGOScientific:2017vwq}. In 2019, the EHT collaboration released the first image of the shadow cast by the supermassive black hole harbored by the elliptical galaxy M87 belonging to the nearby Virgo galaxy cluster \cite{EventHorizonTelescope:2019dse}. These remarkable achievements have opened up new avenues for studying the universe and have further validated the predictions of Einstein's theory of general relativity.

However, there are several areas where (GR) still requires improvement in order to accurately describe various physical phenomena. One such area pertains to the observed acceleration of the universe, as indicated by the seminal works of type Ia supernovae (standard candles) \cite{SupernovaSearchTeam:1998fmf,SupernovaCosmologyProject:1998vns}. This suggests the existence of a new component (dark energy) with negative pressure, which is responsible for this acceleration. Initially, the addition of a cosmological constant to Einstein's field equation was seen as the most plausible solution, despite the discrepancy between the observed value and the theoretical prediction from vacuum energy in QFT \cite{Weinberg:1988cp,Carroll:2000fy,Peebles:2002gy,Frieman:2008sn}. Another avenues consider that dark energy  could be explained in terms of a dynamical field with really a small mass \cite{Martin:2008qp,Tsujikawa:2013fta,Armendariz-Picon:2000nqq}. A modified theory of gravity could also explain the universe's acceleration on a cosmological scale (see \cite{Clifton:2011jh,Joyce:2016vqv,Nojiri:2017ncd},  and \cite{Heisenberg:2018vsk}). A more reliable approach could be introducing a new degree of freedom, a scalar field, coupled to curvature terms without spoiling all the nice properties of Einstein's field equations (see the conditions mentioned in Lovelock's theorem \cite{Lovelock:1972vz}) and without introducing Ostrogradsky instability  \cite{Langlois:2015cwa,deRham:2016wji}.  Among the new healthy theories with all those properties, the so-called Hordensky's theories \cite{Horndeski:1974wa} have attracted considerable attention in recent years \cite{Gleyzes:2014dya,Barausse:2015wia,Kase:2018aps,Kobayashi:2019hrl}. From an observational standpoint, there are compelling reasons to consider theories such as Hordensky's and beyond. These theories have successfully met rigorous constraints based on multi-messenger gravitational-wave astronomy.  The detection of the GW170817 binary neutron star merger, along with its associated electromagnetic counterparts, has provided an accurate bound on the speed of gravitational waves, $|c_{g}/c -1|<10^{-16}$ \cite{Ezquiaga:2017ekz,Creminelli:2017sry}. 

In connection with the frameworks mentioned above, an appealing version of the Einstein-Gauss-Bonnet theory in four dimensions (4D-EGB) was constructed after a regularization process \cite{Fernandes:2020nbq, Hennigar:2020lsl, Lu:2020iav, Kobayashi:2020wqy}. As is well-known, the GB curvature invariant reduces to a topological surface term in four dimensions, so whether the EGB theory can be defined correctly in the limit $D\rightarrow 4$ was an open question \cite{Glavan:2019inb,Gurses:2020rxb,Arrechea:2020gjw,Ai:2020peo,Fernandes:2022zrq}. A neat way to deal with this issue is by considering a conformal transformation and employing a subtraction method to render the theory well-defined in that limit \cite{Fernandes:2022zrq}. The conformal factor involves a scalar field that introduces a new degree of freedom into the model. Interestingly enough,  the regularized 4D-EGB model can be accommodated as a particular case within the shift-symmetric Horndeski's theories \cite{Horndeski:1974wa,Kobayashi:2019hrl}.   Another possibility for obtaining a well-defined 4D-EGB theory can be achieved by implementing a  Kaluza-Klein dimensional reduction from a higher-dimensional Einstein-Gauss-Bonnet gravity plus a scalar field. In that case, the scalar field encodes the information of the volume's size of the compact internal dimension with maximal symmetry \cite{Lu:2020iav}. 

One way to characterize new theories is by using observational tests on different scales, from the solar system to the cosmological scales \cite{PhysRevD.102.084005}. In that regard,  the regularized 4D-EGB model was scrutinized on several fronts. Considering the limits on the GW velocity from the observations of GW170817 and GRB 170817A, the bounds on the rescaled coupling parameter reads $0<\hat{\alpha}<10^{50}\,{\rm{eV}}^{-2}$ \cite{Feng:2020duo}. Stronger limits on positive values come from binary black hole systems, resulting in the overall range $\hat{\alpha}<10^{8}\,{\rm{m}}^{2}$ \cite{PhysRevD.102.084005}.  In the weak-field regime where the gravitational field is considerably tiny,  using the LAGEOS satellite data, one can obtain comparable bounds on the Gauss-Bonnet coupling, $\hat{\alpha}<10^{10}{\rm{m}}^{2}$ \cite{PhysRevD.102.084005}. Besides, assuming that the lightest component of the GW190814 event corresponds to black hole  with mass $M= 2.59^{+0.08}_{-0.09}\,M_{\odot}$, then the bound on GB coupling reduces  several orders of magnitude, $\hat{\alpha}< 59 \times 10^{6}{\rm{m}}^{2}$ \cite{Charmousis:2021npl}.\footnote{The most stringent limit on negative value of GB coupling comes from the condition that atomic nuclei must not be shielded by a horizon, yielding $-\hat{\alpha}> 10^{-83}H^{2}_{0}$ \cite{Charmousis:2021npl}. Besides, primordial GWs also provide another similar constraint on the GB coupling, $|\hat{\alpha}|<10^{36}{\rm{m}}^{2}$ \cite{Aoki:2020iwm},\cite{Fernandes:2022zrq}}

In 2024, new constraints were reported within the context of FRW background with non-zero spatial curvature \cite{Zanoletti:2023ori}. To be more precise, the latter case leads to a modified Friedmann equation for the Hubble parameter after integrating the scalar field dynamical equation; in fact, the non-linear Friedmann equation has a new dark radiation term multiplied by the GB coupling constant. Physically speaking, the extra-dark radiation term will modify the (massless) species number of neutrinos at both the background and perturbative levels \cite{Zanoletti:2023ori}. The constraints based on the  ACTPol data alone implies
$\hat{\alpha}C^{2}= (-9\pm 6)\times 10^{-6}H^{2}_{0}$, where $\hat{\alpha}C^{2}=\Omega^{\rm{extra}}_{\rm{rad}}H^{2}_{0}$ stands for the geometrical dark radiation term \cite{Zanoletti:2023ori}.

We aim to extend previous works on astrophysical constraints on the regularized four-dimensional Einstein-Gauss-Bonnet theory by reevaluating the parametrized post-Newtonian (PPN) formalism \cite{Will:1993ns} within the context of the 4D-EGB theory and adding stronger astrophysical observational constraints along with a whole discussion about bounds comings from complementary probes.

The paper proceeds as follows. We obtain the effective field equations for the metric and scalar fields in Section \ref{sec:feq}. Section \ref{sec:metric} is devoted to the determination of the post-Newtonian form of the metric 
within the PPN formalism standards, where conclusions on the $\gamma$ parameter are already obtained.
We explore in Section \ref{sec:cons} the post-Newtonian conservation laws statements to drawn conclusions on the $\zeta$'s and $\alpha$'s parameters. All those parameters does maintain their physical meaning unaltered and it is shown to have the same value as in GR.
Section \ref{sec:body} is dedicated to computing the center-of-mass acceleration of an extended object in a system of $N$-bodies within the PN approximation, where we capture the standard GR term plus additional terms proportional to the Gauss-Bonnet coupling, $\a$. With the previous results, we can estimate the shift in the orbital periastron for a binary system due to the GB coupling, which is done in Section \ref{sec:periastron}. We also examine the periastron advance rate in the cases of Mercury and the double pulsar. Section \ref{sec:beta} discusses the role played by the $\beta$ parameter in the PPN approach and the Nordtvedt effect, along with an analysis on haw to extend th formalism to include the corrections of the EGB coupling. We state our conclusions in Section \ref{sec:conclusion}.

\section{The regularized 4D-EGB theory}\label{sec:feq}

In this section, we will derive the field equations of the regularized 4D-EGB theory as obtained in the previous Refs.\cite{Fernandes:2020nbq, Hennigar:2020lsl, Lu:2020iav, Kobayashi:2020wqy}.
Our starting point is to consider  the gravity sector of the EGB without cosmological constant in   $D$ dimensions,
\begin{equation}
S=\int_{\mathcal{M}} d^D x \frac{\sqrt{-g}}{2\kappa} \left( R + \alpha \mathcal{G} \right) \ ,
\label{EGBactionD}
\end{equation}
where $\kappa=8\pi G/c^4$, with $G$ the gravitational constant and $c$ the velocity of light in vacuum whereas  $\alpha$ stands for a coupling constant associated to the quadratic term, $\mathcal{G}$. The regularization process involves a conformal transformation, followed by a redefinition of the coupling constant. Specifically, we apply $\Tilde{g}_{\mu \nu} = e^{2 \Phi} g_{\mu \nu}$, which then transforms the square root of the determinant as $\sqrt{-\Tilde{g}} = e^{D \Phi} \sqrt{-g}$.  The Gauss-Bonnet term changes as follows \cite{Fernandes:2022zrq},
\begin{widetext}
\begin{align}
    \notag \sqrt{-\Tilde{g}}\Tilde{\mathcal{G}} &= e^{(D-4) \Phi} \sqrt{-g}\Big[\mathcal{G}-8(D-3)R^{\mu\nu}(\nabla_{\mu}\Phi\nabla_{\nu}\Phi-\nabla_{\mu}\nabla_{\nu}\Phi)-2(D-3)(D-4)R (\nabla \Phi)^{2} \\ \notag
    &+4(D-2)(D-3)^{2}\Box\Phi(\nabla \Phi)^{2}-4(D-2)(D-3)\nabla_{\mu}\nabla_{\nu}\Phi\nabla^{\mu}\nabla^{\nu}\Phi -4(D-3)R\Box\Phi\\ 
    &+4(D-2)(D-3)(\Box \Phi)^{2}+8(D-3)(\nabla_{\mu}\Phi\nabla_{\nu}\Phi)(\nabla_{\mu}\nabla_{\nu}\Phi)+(D-1)(D-2)(D-3)(D-4)(\nabla \Phi)^{4}
    \Big],
\end{align}
\end{widetext}
  The regularization method requires to subtract a counter-term $ \alpha \sqrt{-\Tilde g} \Tilde{\mathcal{G}}$ in the original action.  The idea is to expand the exponential around $D=4$ and keep the lowest order in $(D-4)$ and neglect all higher order terms, namely,  $e^{(D-4) \Phi}=1+(D-4)\Phi +\mathcal{O}[(D-4)^2]$. We use  the Bianchi identities  along with the Bochner's formula in curved spacetime, $\frac{1}{2} \nabla^\mu \left(\nabla \Phi \right)^2  = \left(\nabla_\mu \nabla_\nu \Phi\right)^2 +\nabla_\mu \Big( \Box \Phi \nabla^\mu \Phi\Big)-\left( \Box \Phi \right)^2  + R^{\mu \nu} \nabla_\mu \Phi \nabla_\nu \Phi$, for simplifying the action. In doing so, 
we integrate by parts, discarding any surface terms. We  set $D\rightarrow 4$ for the terms that are not of the form $D-4$, and the divergences in the original action is canceled provided the term proportional to $(D-4)$ remains finite as they are multiplied by the re-scaled coupling $\hat{\alpha}/(D-4)$ \cite{Fernandes:2020nbq, Hennigar:2020lsl, Lu:2020iav, Kobayashi:2020wqy}. The resulting action of the system reads \cite{Fernandes:2020nbq, Hennigar:2020lsl, Lu:2020iav, Kobayashi:2020wqy, Easson:2020mpq},
\begin{align}
	S^{\text{4D-EGB}}_{\text{reg}}=\int d^{4} x \frac{\sqrt{-g}}{2\kappa}\big[&R+\hat{\alpha}\big(4 G^{\mu \nu} \nabla_{\mu} \Phi \nabla_{\nu} \Phi-\Phi \mathcal{G} \nonumber \\
	&+4 \square \Phi(\nabla \Phi)^{2}+2(\nabla \Phi)^{4}\big)\big]+S_{m},
	\label{eq:action}
\end{align}
where we identify $G_{\mu\nu}$ as the Einstein tensor whereas ${\cal G}=R_{\alpha \beta \mu \nu} R^{\alpha \beta \mu \nu} - 4 R_{\mu \nu} R^{\mu \nu} +R^2$ denotes the Gauss-Bonnet term. As usual,  $R$ refers to the Ricci scalar, $S_m$ describes the matter action, and $\nabla_{\mu}$ indicates a covariant derivative. The extra degree of freedom,$\Phi$,   comes from the scalar field that encodes the conformal transformations that allow us to regularize the gravitational action for $D=4$ dimensions \cite{Fernandes:2020nbq}. For an entire debate on the nature of this regularization process, we refer the reader to the following works \cite{Fernandes:2020nbq, Hennigar:2020lsl, Lu:2020iav, Kobayashi:2020wqy, Easson:2020mpq,  Mahapatra:2020rds, Bonifacio:2020vbk}.
\begin{widetext}
 To explore the new modifications introduced by this regularization process, we need to obtain the field's equation of the regularized 4D-EGB theory \eqref{eq:action}. We carry on by computing the variation of action \eqref{eq:action} with respect to the metric, it is given by 
\begin{equation} \label{feq-metric}
	G_{\mu \nu} =  \hat{\alpha} \hat{\mathcal{H}}_{\mu \nu} +\kappa T_{\mu \nu} \, ,
\end{equation}
where $T_{\mu \nu}$ is the energy-momentum tensor associated with the matter fields and the tensor $\hat{\mathcal{H}}_{\mu \nu}$ can be  written as
	\begin{equation} \label{H}
		\begin{aligned}
			\H_{\mu\nu} =&  \  2R(\nabla_\mu \nabla_\nu \Phi - \nabla_\mu\Phi \nabla_\nu \Phi) + 2G_{\mu \nu}\left(\dpp-2\dal\right) + 4G_{\nu \alpha} \left(\nabla^\alpha \nabla_\mu \Phi -\nabla^\alpha \Phi \nabla_\mu \Phi\right)\\
			& + 4G_{\mu \alpha} \left(\nabla^\alpha \nabla_\nu \Phi - \nabla^\alpha \Phi \nabla_\nu \Phi\right) + 4R_{\mu \alpha \nu \beta}\left(\nabla^\beta \nabla^\alpha \Phi - \nabla^\alpha \Phi \nabla^\beta \Phi\right)+ 4\nabla_\alpha\nabla_\nu \Phi \left(\nabla^\alpha \Phi \nabla_\mu \Phi - \nabla^\alpha \nabla_\mu \Phi \right)\\
			& +4 \nabla_\alpha \nabla_\mu \Phi \nabla^\alpha \Phi \nabla_\nu \Phi - 4\nabla_\mu \Phi \nabla_\nu \Phi \left(\dpp + \dal\right)+4\dal \nabla_\nu \nabla_\mu \Phi - g_{\mu \nu} \Big( 2R\left(\dal - \dpp \right)\\
			& + 4 G^{\alpha \beta} \left( \nabla_\beta \nabla_\alpha \Phi - \nabla_\alpha \Phi \nabla_\beta \Phi \right) + 2(\dal)^2 - \left( \nabla \Phi\right)^4 + 2\nabla_\beta \nabla_\alpha\Phi\left(2\nabla^\alpha \Phi \nabla^\beta \Phi - \nabla^\beta \nabla^\alpha \Phi \right)
			\Big).
		\end{aligned}
	\end{equation}
The variation of the action \eqref{eq:action} with respect to the scalar field, $\Phi$, leads to its  equation of motion, 
	\begin{equation} \label{feq-phi}
		\begin{aligned}
			&R^{\mu \nu} \nabla_{\mu} \Phi \nabla_{\nu} \Phi - G^{\mu \nu}\nabla_\mu \nabla_\nu \Phi - \dal \dpp +\nabla_\mu \nabla_\nu\Phi \nabla^\mu \nabla^\nu\Phi 
			- (\dal)^2 - 2\nabla_\mu \Phi \nabla_\nu \Phi \nabla^\mu \nabla^\nu \Phi = \frac{1}{8}\mathcal{G}.
		\end{aligned}
	\end{equation}
\end{widetext}

The field equation \eqref{feq-metric} can be rearranged with the help of its trace by writing the Ricci scalar in terms of $T$ and $\H$, i.e., the respective traces of the energy-momentum tensor and the tensor $\H_{\mu\nu}$. This leads us to recast the metric field equations in a more suitable way for solving them within the post-Newtonian approximation as the standard Einstein's field equation plus a new contribution encoded in the $\H_{\mu\nu}$ tensor,
\begin{equation}\label{feq}
R_{\mu\nu}=8\pi\left(T_{\mu\nu}-\frac{T}{2}\,g_{\mu\nu}\right)+ \a\left(\H_{\mu\nu}-\frac{\H}{2}\,g_{\mu\nu}\right),
\end{equation}
where we chose a geometrodynamics system of units by selecting $G=c=1$. The aforesaid theory \eqref{feq}  is a particular case of a broader class of shift-symmetric Horndeski's theories. These theories have the property that their equations of motion do not change when a constant shifts the scalar field, $\Phi \rightarrow \Phi + c$, where $c$ is a constant; therefore, the theory admits a conserved Noether current \cite{Horndeski:1974wa,Kobayashi:2019hrl}. Specifically, it corresponds to $G_2=8 \hat{\alpha} X^2$, $G_3=8 \hat{\alpha} X$, $G_4=1+4 \hat{\alpha} X$, and $G_5 = 4 \hat{\alpha} \ln X$ (where $X=-\frac{1}{2} \nabla_{\mu} \Phi \nabla^{\mu} \Phi$). General Horndeski theories have been previously analyzed within the post-Newtonian approximation scheme, but only by considering models in which the $G_n$ functions can be expressed  as a Taylor series around  $X=0$; this is not the case for $G_5$  \cite{Hohmann:2015kra}. Notwithstanding, equations \eqref{feq-metric} and \eqref{feq-phi} can be regularly treated within the post-Newtonian limit, as seen in the following sections.

\section{Post-Newtonian expansion}\label{sec:metric}

Let us focus on solving the theory's field equation in a regime where the gravitational field is weak everywhere and matter moves in small velocities compared to light speed.  We proceed by expanding equations \eqref{feq} and \eqref{feq-phi} using the following perturbation's scheme,
\begin{align}\label{metricexp}
g_{\mu\nu}&=\eta_{\mu\nu}+h_{\mu\nu}\,,\\
\Phi&=\phi_0+\phi,
\end{align}
where $h_{\mu\nu}$ encodes the metric perturbations concerning Minkowski's background, whereas $\phi$ stands for the scalar field perturbation around its constant background value, $\phi_0$.  

For the matter content, we use a perfect fluid approach,
\begin{align}
	T^{\mu\nu}=\left(\rho+\rho\Pi+p\right)u^{\mu}u^{\nu}+pg^{\mu\nu}\,,
\end{align}
where $\rho$ is the mass density, $\Pi$ is the fluid's internal energy per unity mass, $p$ is the pressure and $u^{\mu}=u^{0}(1,\bs v)$ is the four-velocity of the fluid. In the post-Newtonian approximation scheme, the energy-momentum tensor components should be expanded in power of $v$, taking into account also that $p/\rho\sim\Pi\sim O(2)$. Here and after, we use the notation $O(N)$ to represent quantities of order $v^N$.Moreover, time derivatives must be considered as order of magnitude one, $\p_t\sim O(1)$,  since the dynamical time scale in the solar system is governed by the motion of planets.

As expected, the metric perturbations $h_{\mu\nu}$ will also be expanded in powers of $v$. To obtain the first post-Newtonian corrections to the equations of motion of massive bodies, one needs to know $h_{00}$ up to order $v^{4}$, $h_{0i}$ up to order $v^{3}$ and $h_{ij}$ up to order $v^{2}$, where Latin indices run over the three spatial dimensions. The approximation procedure described above is standard in the post-Newtonian approach and is explained in great detail in Refs. \cite{Will:1993ns,PoissonWill}, which we will follow closely.

The fluid dynamics is subjected not only to conservation of the energy-momentum tensor, $\nabla_{\nu}T^{\mu\nu}=0$, but also to the conservation of rest mass density, 
\begin{equation}\label{consmatter}
\nabla_{\mu}(\rho u^{\mu})=0 \, .
\end{equation}
This equation can be re-expressed as an effective flat-space continuity equation as follows,
\begin{equation}\label{cont}
\p_t(\rs)+\p_i(\rs v^{i})=0,
\end{equation}
where the conserved density is defined in terms of zero component of the four-velocity field and the square root of the determinant,
\begin{equation}\label{rstar}
\rs\equiv {u^{0}}\sqrt{-g}\,\rho.
\end{equation}
Since $\rs$ is the conserved density, in the sense of \eqref{cont}, it is more convenient to use it to express the energy-momentum tensor components. As will become clear later, having the gravitational potentials expressed in terms of the conserved density will lead to an easy way to integrate the equations of motion of massive bodies.

Applying the above approximations, one can verify that the first non-vanishing terms in equation \eqref{feq-phi} and for $\H_{\mu\nu}$ occur in the fourth order only. Consequently, 4D-EGB gravity will deviates from GR only at the fourth order [see Eq. \eqref{feq}].  Thus, up to $O(3)$, the metric components reads,
\begin{align}
g_{00}&= -1 + 2U +O(4),\label{g002}\\
g_{0i}&=-4\,V^{i} + O(5),\label{g0i3}\\
g_{ij}&=(1+2U)\delta_{ij} + O(4),\label{gij2}
\end{align}
where $U$ is the negative of the Newtonian potential,
\begin{equation}
U= \int\frac{\rs(t,\bs x') \,d^{3}x'}{|\bs x-\bs x'|},
\end{equation}
and,
\begin{equation}\label{vi}
V^i= \int\frac{\rs(t,\bs x')v^{i}}{|\bs x-\bs x'|}\,d^{3}x'.
\end{equation}

To determine the fourth-order term of $g_{00}$ we first consider equation \eqref{feq-phi}, which gives,
\begin{equation}
\p_{ij}\phi \,  \p^{ij}\phi - \left( \nabla^2 \phi \right)^2 = \p_{ij}U \,  \p^{ij}U - \left( \nabla^2 U \right)^2,
\end{equation}
where we are using the notation $\p_{ij}=\p/\p x^i\p x^j$. The simplest solution to the above equations reads,
\begin{equation} \label{phi}
\phi = \pm \,U  .
\end{equation}
Using Eqs. \eqref{g002}-\eqref{gij2} and \eqref{phi},  the $00$-component of field equation \eqref{feq} becomes,
\begin{align}
\nabla^{2}h_{00}= -\rs \left(1+\frac{3}{2}v^{2} -U +\Pi +3p \right)\nonumber\\
  -2\p_{tt}U  \, \pm \, \a\,\G+O(6),\label{nablah00}
\end{align}
where $\G$ is the fourth-order approximation of Gauss-Bonnet invariant,
\begin{equation}
	\G=8\,\p_{ij}U \,  \p^{ij}U - 8\left( \nabla^2 U \right)^2.
\end{equation}
Assuming asymptotically flatness, the general solution of \eqref{nablah00} reads,
\begin{equation}\label{h004}
h_{00}=-2U +2(\Psi -U^{2} \mp \a\,\PhiG) +O(6),
\end{equation}
with
\begin{equation}\label{psi}
\Psi\equiv\dfrac{3}{2}\,\Phi_1-\Phi_2+\Phi_3+3\Phi_4 +\frac{1}{2}\p_{tt}X,\\[1ex]
\end{equation}
and the PN  potentials that appear can be recast as follows,
\begin{align}
\Phi_1=\int\dfrac{\rs{}' v'^{2}}{|\bs x- \bs x'|}\,d^{3}x',\quad \Phi_2=\int\dfrac{\rs{}' U'}{|\bs x- \bs x'|}\,d^{3}x',\\[1ex]
\Phi_3=\int\dfrac{\rs{}' \Pi'}{|\bs x- \bs x'|}\,d^{3}x',\quad \Phi_4=\int\dfrac{p'}{|\bs x- \bs x'|}\,d^{3}x' \, ,\\[1ex]
X=\int\rs{}'|\bs x- \bs x'|\,d^{3}x',\quad \PhiG=\frac{1}{4\pi}\int\dfrac{\G\,d^{3}x' }{|\bs x- \bs x'|}\,.
\end{align}
The prime symbol indicates that a  quantity is evaluated at the spatial coordinates $x'$.

In brief, the post-Newtonian metric of 4D-EGB theory is given by
\begin{align}
g_{00}&= -1 + 2U +2(\Psi -U^{2}) \mp \a\,\PhiG +O(6),\label{g00}\\[1ex]
g_{0i}&=-4\,V^{i} + O(5),\label{g0i}\\[1ex]
g_{ij}&=(1+2U)\delta_{ij} + O(4).\label{gij}
\end{align}
The above metric is consistent with previous analyses on weak-field and slow-motion approximations of 4D-EGB gravity \cite{PhysRevD.102.084005}. However, several differences deserve to be mentioned:  $i)$ we are working with harmonic gauge instead of the traditional PPN gauge, and $ii)$ we are defining potentials in terms of the conserved density. Notwithstanding, this does not change the form of the correction brought by this theory, represented by the potential $\PhiG$. This potential is not covered by the PPN formalism, implying that it is not correct to read off the PPN parameters directly from the metric structure. One must investigate the equations of motion of massive bodies and light rays to correctly identify the new potential influence in the standard PPN parameters.

It is worth to mention this is not a particularity of 4D-EGB model, once several modern theories of gravity does not fit entirely to the PPN formalism assumptions. It seems not to exist a fundamental rule that explains or determines when a model may include extra-PPN potentials on it. An a priori simple theory, like purely scalar gravity, can produces an extremely knotty PN metric \cite{Bittencourt:2016smd}, while GR extensions motivated by nontrivial renormalization group effects at large scales may not spoil the PPN framework at all \cite{Toniato:2017wmk}. 
Determinations of PPN parameters for theories with extra-PPN potentials is usually possible, as in the case of Palatini $f(R)$ gravity \cite{Toniato:2019rrd}, for instance. But sometimes some extra although reasonable assumptions are needed, as in scalar-tensor and metric $f(R)$ formulations \cite{Toniato:2021vmt, alves2024}. Fortunately, the latter is not the case of 4D-EGB.

\subsection{The \texorpdfstring{$\boldsymbol{\gamma}$}{g} parameter}

For the propagation of light, the first post-Newtonian correction is obtained by considering only the second-order terms of the metric. The PPN formalism only assumes the Newtonian potential to be present at order two, and any deviation from GR is encoded in the single parameter $\gamma$ multiplying the $g_{ii}$ components (see Ref. \cite{Toniato:2021vmt} for a detailed discussion on the  parametrization of propagation of light). Thus, once the 4D-EGB theory produces only the Newtonian potential at the second-order metric, one can conclude that the physical meaning of the $\gamma$ parameter remains unaltered. Since the spatial component of the second-order metric is the same as in GR [eq. \eqref{gij}], it is obtained
\begin{equation}
    \gamma=1,
\end{equation}
and a full agreement with tests in the solar system based on propagation of light. The remaining PPN parameters can be inferred from the calculations of the conserved quantities and the equations of motion of massive bodies. In the following sections, we will perform this analysis.

\section{Conserved Quantities}\label{sec:cons}

In this section, we obtain the expressions for a fluid system's total mass-energy and momentum, which are conserved in the post-Newtonian limit. The re-scaled density $\rs$ allows one to define the fluid material mass in a given volume as follows,
\begin{equation}\label{mass}
m=\int\rs d^{3}x \, .
\end{equation}
Hence, for an isolated body, by using the continuity equation for $\rho^*$ [cf. eq. \eqref{cont}], it is shown that mass $m$ is conserved, i.e.
\begin{equation}\label{mcons}
\dfrac{dm}{dt}=0\, .
\end{equation} 
The above result assumes that density $\rho^*$ is zero at the boundary of the integration volume. The statement above only depends on the fluid definition, not on the used gravity model. The situation is different for energy and momentum conservation.

From the energy-momentum tensor conservation, we write
\begin{equation}\label{divt}
\nabla_{\nu}T^{\mu\nu}=\p_\nu(\sqrt{-g}\,T^{\mu\nu})+ \Gamma^{\mu}_{\lambda\nu}(\sqrt{-g}\,T^{\lambda\nu})=0.
\end{equation}
We want to integrate the above expression in its post-Newtonian approximation and obtain the conserved quantities in 4D-EGB gravity. For practical reasons, one can work with a redefinition of eq. \eqref{g00} such as,
\begin{equation}
\label{g00red}
 g_{00}=(g_{00})_{\scriptscriptstyle\rm GR} \mp \a\,\PhiG/2,
\end{equation}
where $(g_{00})_{\scriptscriptstyle\rm GR}$ stands for  the post-Newtonian metric from GR. Then, one can easily separate the contributions inserted by the 4D-EGB theory and use the standard GR results as shown in Ref. \cite{PoissonWill} for instance.

The statement of energy conservation is derived from the $\mu=0$ component of \eqref{divt}, where there is no fourth-order metric contributions. Hence the results are the same as in GR: the leading order reproduces the continuity equation \eqref{cont} and the next order gives the energy which is conserved at post-Newtonian regime, namely
\begin{equation}\label{energy}
E\equiv\int\left(\frac{1}{2}\rs v^{2}  + \rs\Pi - \frac{1}{2}\rs U \right)d^{3}x.
\end{equation}
Thus, the total mass-energy is given by \begin{equation}
M=m+E.\label{mass-energy}
\end{equation}

For the conservation of the total momentum, we consider only the case $\mu = i$ in eq.~\eqref{divt}, which can be written as follows,
\begin{equation}
	(\nabla_{\nu}T^{\nu}_{~i})_{\s\rm GR} \pm \dfrac{\a}{2}\,\rs \nabla_i\PhiG=0,\label{mu-i}
\end{equation} 
where $(\nabla_{\nu}T^{\nu}_{~i})_{\s\rm GR}$ stands, once again, for the ordinary GR terms. By integrating this expression over the volume occupied by the fluid, we are able to derive the total momentum which is conserved. We only need to integrate the last term in \eqref{mu-i} since the results from the GR contributions are already known \cite{PoissonWill}. To do so, we first verify that (see Ref.\cite{PhysRevD.102.084005}) $\PhiG$ can be expressed in terms of $U$ and $\Phi_7$ as follows,
\begin{equation}\label{phig}
\PhiG = -8 \left( \frac{1}{2} \vert \nabla U \vert^2 - \Phi_7 \right) \, ,
\end{equation}
where we have defined,\footnote{Symbols $\Phi_5$ and $\Phi_6$ are already used within GR to represent other gravitational potentials. This is the reason why we use $\Phi_7$ to represent this new potential. It is the same potential described in Ref. \cite{PhysRevD.102.084005} as $\psi_1$.}
\begin{equation}
\Phi_7 = -\int \rs{}'\p'_jU'\frac{({x} - {x}' )^j}{\vert {\bf x} - {\bf x}' \vert^3 } d^3 x' .
\end{equation}

By integrating the gradient of $\Phi_7$ and using  the identity $\p_i f(x-x')=-\p_i'f(x-x')$ plus an exchange of variables, $x\leftrightarrow x'$, one arrives at the following expression,
\begin{align}
	\int\rs\p_i\Phi_7\,d^3x&=\int\rs\p^jU\p_{ij}U\,d^3x,\nonumber\\[1ex]
	&=\frac{1}{2}\int\rs\p_i(|\nabla U|^2)d^3x.\label{dphi7}
\end{align}
Combining \eqref{phig} and \eqref{dphi7}, one shows that the following relation holds true,
\begin{equation}
	\int\rs\p_i\PhiG\,d^3x=0.
\end{equation}
Consequently,  after integrating \eqref{mu-i}, one gets the same results as in GR for the total conserved momentum,
\begin{align}
P^{j}=&\int\rs v^{j}\bigg(1 + \frac{v^{2}}{2} - \frac{U}{2} + \Pi + \frac{3p}{\rs}\bigg)d^{3}x \ \notag \\[1ex]
&- \frac{1}{2}\int\rs W^{j}d^{3}x, \label{momentum}
\end{align}
where the post-Newtonian gravitational potential, $W^i$,  is written  as 
\begin{equation}
	W^i = \int \frac{\rs \, [{\bf v \cdot (x - x')}] (x-x')^i}{\vert {\bf x} - {\bf x'} \vert^3} d^3 x' \, .
\end{equation}

The previous results show that 4D-EGB gravity does not violate the total conservation of energy and momentum in the PN regime. This is an expected outcome since the theory is Lagrangian based and, as shown in Ref. \cite{Lee:1974nq}, they should not violate PN conservation laws. 

\subsection{The \texorpdfstring{$\boldsymbol{\zeta}$'s and $\boldsymbol{\alpha}$'s}{z's and a's} parameters}
In the context of the PPN formalism, the results obtained here directly show that the PPN parameters $\zeta_1,\zeta_2,\zeta_3,\zeta_4$ and $\alpha_3$ do not have their physical meaning changed by the presence of the potential $\PhiG$ in the PN expansion. Their values, if different from zero, continue to indicate a violation of energy and momentum conservation. Thus, one can conclude that, for 4D-EGB theory, one has
\begin{equation}\label{semiconservative}
\zeta_1=\zeta_2=\zeta_3=\zeta_4=\alpha_3=0.
\end{equation}
Since the expressions \eqref{energy} and \eqref{momentum} are identical to  GR counterparts, one can also conclude that parameters $\alpha_1$ and $\alpha_2$ retain their physical meaning unchanged since no extra-PPN potentials are present in the expressions for the conserved energy and momentum. Moreover, a direct comparison with the general PPN expression for $P^j$ assuming results in \eqref{semiconservative}, the following is obtained,
\begin{equation}
    \alpha_1=\alpha_2=0.
\end{equation}
The latter result shows that 4D-EGB theory does not have any preferred frame effects (see Appendix \ref{appendix} for details).

\section{Equation of motion of massive bodies}\label{sec:body}

In this section, we split the fluid description of the source into $N$ separated bodies. We aim to obtain the PN equations of motion for the body's center-of-mass positions. This is a realistic way to deal with the trajectories of massive and finite-volume bodies instead of assuming that they can be treated as test particles. Each body indexed by $A$ has a material mass given by
\begin{equation}\label{massa}
m_A=\int_{A}\rs d^{3}x.
\end{equation}
The volume where the integration above is calculated is a time-independent region of space that extends beyond the volume occupied by the body. It is large enough that, in a time interval $dt$, the body does not cross its boundary surface, but it is also small enough not to intersect with any other body of the system. The center-of-mass of a body $A$, its velocity, and acceleration are then defined as
\begin{align}
\bs{r}_A(t)\equiv& \ \dfrac{1}{m_A}\int\rs\bs{x}\,d^{3}x,\\[1ex]
\bs{v}_A(t)\equiv& \ \dfrac{d\bs{r}_A}{dt}=\dfrac{1}{m_A}\int\rs\bs{v}\,d^{3}x,\\[1ex]
\bs{a}_A(t)\equiv& \ \dfrac{d\bs{v}_A}{dt}=\dfrac{1}{m_A}\int\rs\dfrac{d\bs{v}}{dt}\,d^{3}x \,.  \label{a} 
\end{align}
The integrand of eq.~\eqref{a} is found from the Euler equation with its PN extension, which can be derived from eq.~\eqref{mu-i}. Each integral is then interpreted as a force that can be calculated according to the techniques depicted in Ref. \cite{PoissonWill}. Ordinary GR terms give rise to 18 forces, while the 4D-EGB correction can be split into two more contributions, namely,
\begin{align}
	F^i_{\s 19}&=\mp\, 4\a \int_A\rs \p_{j}U\p_{ij}U\,d^3x,\\[1ex]
	F^i_{\s 20}&= \pm\, 4\a \int_A\rs \p^i\Phi_7\,d^3x.
\end{align}
These forces do not cancel one with each other, as in the previous section, since the potentials in the integrand are integrals over the volume occupied by the fluid, while the forces are integrals over the volume surrounding the body $A$ only.

To explicitly calculate the forces above, we need to separate the potentials into an internal part, produced by the body $A$, and an external one sourced by the remaining bodies of the system. For instance, $U$  can be written as follows,
\begin{align}
U&=\int_{A}\dfrac{\rs{}'}{|\bs x-\bs x'|}\,d^{3}x' + \sum_{B\neq A}\int_B\dfrac{\rs{}'}{|\bs x-\bs x'|}\,d^{3}x',\notag\\[1ex]
&\equiv U_{\s A} + U_{\s A}^{\s ext}.\label{ext}
\end{align}
Assuming a wide separation between bodies implies that, when evaluating an external potential within the body $A$, it can be expanded in a Taylor series. As an example, one has
\begin{equation}
U_{A}^{ext}(t,\bs x)\approx U_{A}^{ext}(t,\bs r_A)+ \bx^{j}\p_jU_{A}^{ext}(t,\bs r_A)  +\dots\,,\label{exp}
\end{equation}
where $\bs\bx=\bs x- \bs r_A$.
The series is truncated in its second term since the next terms are at least of order ${\cal{O}}(R_A/r_{AB})^{2}$, where $R_A$ is the typical body radius and $r_{AB}=|\bs r_A -\bs r_B|$ is the interbody distance. This expansion is used to extract the external pieces of potentials from the integrals. Moreover, each body of the system is assumed to be reflection-symmetric above its own center of mass, i.e., $\rs(t,\bs r_A-\bs\bx)=\rs(t,\bs r_A+\bs\bx)$. This symmetry allows us to eliminate any integral having an odd number of internal vectors, such as $\bs\bx$.

Starting with $F^i_{\s 20}$, one has
\begin{equation}
F^i_{\s 20}= \pm\, 4\a \int\rs\p^i\Phi_{\s 7,A}d^3x \pm\, 4\a m_A\p^i\Phi_7^{\s ext}(t,\bs r_A).
\end{equation}
Using the brief notation $s=|\bs x - \bs x'|$ one can write $\p_i\Phi_{\s 7,A}=\int_{\s A}\rs{}'\p_j'U'\p_{ij}s^{-1}\,d^3x'$ and, after separating the potential inside the integrand in its internal and external parts, one obtains,
\begin{align}
	\int\rs\p_i\Phi_{\s 7,A}d^3x=& 	\int_{\s A}\rs\p_jU_{\s A}\p_{ij}U_{\s A}d^3\bx \notag\\
	& \ + \ \p_jU^{\s ext}_{\s A}\int_{\s A}\rs\p_{ij}U_{\s A}\, d^3\bx.\label{f20}
\end{align}
In the above expression, we have changed the variables  $\bs x$ by $\bs\bx$, and used, when necessary, the reflection-symmetric properties discussed before.

For $F^i_{\s 19}$, one has
\begin{align}
	F^i_{\s 19}&=\int_{\s A}\rs\p_jU_{\s A}\p_{ij}U_{\s A}d^3\bx +\p_jU^{\s ext}_{\s A}\int_{\s A}\rs\p_{ij}U_{\s A}\, d^3\bx \notag\\
	& +\p_{ij}U^{\s ext}_{\s A}\int_{\s A}\rs\p_{j}U_{\s A}\, d^3\bx +m_{\s A}\p_{j}U^{\s ext}_{\s A}\p_{ij}U^{\s ext}_{\s A}.
\end{align}
We then note that the first line above is equal to \eqref{f20} and will cancel when both forces are added. The third integral in the r.h.s. above vanishes since it has an odd number of internal vectors $\bx$ and $\bx'$ due to the gradient of $U$. At the end, one has,
\begin{equation}\label{force}
	F^i_{\s 19} + F^i_{\s 20}= \pm\,4\a m_{\s A} \left(\p^i\Phi^{\s ext}_{7,{\s A}} - \p_{j}U^{\s ext}_{\s A}\p_{ij}U^{\s ext}_{\s A}\right).
\end{equation}
Here, the external potentials are always calculated at $\bs x=\bs r_A$ after the derivatives are done.

Now, we have to express the external potentials in terms of the center-of-mass positions. We start by calculating $\p_{j}U^{\s ext}_{\s A}$ and $\p_{ij}U^{\s ext}_{\s A}$. These potentials depend on derivatives of $s$ and integrations over the volume occupied by the bodies $B\neq A$, for instance $\p_{j}U^{\s ext}_{\s A}=\sum_{\s B\neq A}\int\rs{}'\p_js^{-1}d^3x'$. TThus, due to the wide separation between bodies in the system, we can write $s=|\bs x-\bs r_{\s B}- \bs\bx'|$ and expand around $\bs\bx=0$  in order to  obtain the following derivatives,
\begin{align}
	\p_js^{-1}&\approx - \frac{s^j_{\s B}}{s_{\s B}^3}+ \bx'^k\left(\frac{s^j_{\s B}s^k_{\s B}}{s_{\s B}^5}-\dfrac{\delta^{jk}}{s_{\s B}^3}\right),\label{ds}\\[1ex]
	\p_{ij}s^{-1}&\approx - \frac{\delta_{ij}}{s_{\s B}^3} +3\dfrac{s^i_{\s B}s^j_{\s B}}{s^5_{\s B}}\notag\\
	&\quad +3\bx'^k\left(\dfrac{5s^i_{\s B}s^j_{\s B}s^k_{\s B}}{s^7_{\s B}}- 
	\dfrac{\delta^{k(i}s^{j)}_{\s B} +\delta^{ij}s^k_{\s B}}{s^5_{\s B}}	\right), \label{dds}
\end{align}
where we used the short notation $\bs s_{\s B}=\bs x-\bs r_{\s B}$ with $s_{\s B}$ being its norm. Also, indices between parenthesis indicate their symmetric counterpart. With the expansions above at hand, it is easy to verify that,
\begin{align}
	\p_{j}U^{\s ext}_{\s A}(t,\bs r_{\s A})&\approx -\sum_{B\neq A}\dfrac{m_{\s B}n^j_{\s AB}}{r_{\s AB}^2} \\[1ex]
	\p_{ij}U^{\s ext}_{\s A}(t,\bs r_{\s A})&\approx - \sum_{B\neq A}\frac{m_{\s B}}{r_{\s AB}^3}\left(\delta_{ij}- 3n^i_{\s AB} n^j_{\s AB}\right),
\end{align}
where $n^i_{\s AB}=r^i_{\s AB}/r_{\s AB}$. Using the previous above results,  one can write,
\begin{align}\label{f20-final}
\p_{j}U^{\s ext}_{\s A}&\p_{ij}U^{\s ext}_{\s A}=\sum_{B\neq A}\Bigg[-\dfrac{2m_{\s B}^2}{r_{\s AB}^5}\,n_{\s AB}^i \notag\\
&+ \!\!\! \sum_{C\neq A,B}\dfrac{m_{\s B}m_{\s C}}{r_{\s AB}^2r_{\s AC}^3}\big(n_{\s AB}^i-3(\bs n_{\s AB}\cdot\bs n_{\s AC})n_{\s AC}^i\big)\Bigg].
\end{align}

The case of $\p^i\Phi^{\s ext}_{7,{\s A}}$ is more involved since it depends on another potential. This last must be separated into a part generated by body $A$, another by the body $B$ (the same as the integral limit), and a third one sourced by the remaining bodies of the system. By doing that, one can write,
\begin{align}
	\p_j\Phi^{\s ext}_{7,{\s A}}=&\sum_{B\neq A}\bigg[ \int_{\s B}\rs{}'\p'_iU'_{\s A}\p_{ij}s\,d^3x' + \int_{\s B}\rs{}'\p'_iU'_{\s B}\p_{ij}s\,d^3x' \notag\\[1ex]
	&\quad \qquad +\sum_{C\neq A,B}\int_{\s B}\rs{}'\p'_iU'_{\s C}\p_{ij}s\,d^3x' \ \bigg].\label{terms}
\end{align}
The first integral reads,
\begin{equation}
	\int_{\s B}\rs{}'\p'_iU'_{\s A}\p_{ij}s\,d^3x=\!\int_{\s\! A}\!\int_{\s \! B}\!\!\rs{}'\rs{}''\p'_is'\p_{ij}s\,d^3x'd^3x'',\label{int}
\end{equation}
with $s'=|\bs x'- \bs x''|$. *The expansion of $\p_{ij}s$ will be equivalent to \eqref{dds}But the expansion of $\p'_is'$ is not straightforward since $\bs x'$ is expanded around $\bs r_{\s B}$ whereas $\bs x''$ is expanded around $\bs r_{\s A}$. We only need  to get the leading order of this expansion, which gives $\p'_is'\approx n^i_{\s AB}/r_{\s AB}^2$, and the first term in the r.h.s. of \eqref{terms} can be accommodated as,
\begin{equation}
	\sum_{B\neq A}\dfrac{2m_{\s A}m_{\s B}}{r^5_{\s AB}}n^j_{\s AB}.
\end{equation}
The second integral in \eqref{terms} is similar to \eqref{int}, but now,  both limits of integration refer to the volume occupied by the body $B$. Thus, we simply expand $\p_{ij}s$ provided $\p'_is'=(\bx'-\bx'')^i/|\bx'-\bx''|^3$. Using \eqref{dds} once again, one arrive at the following integral,
\begin{align}
	\int_{\s B}\rs{}'\p'_iU'_{\s B}\p_{ij}s\,&d^3x'\approx 3\left(\dfrac{5s^i_{\s B}s^j_{\s B}s^k_{\s B}}{s^7_{\s B}}- 
	\dfrac{\delta^{k(i}s^{j)}_{\s B} +\delta^{ij}s^k_{\s B}}{s^5_{\s B}}	\right)\notag\\
	&\times \int_{\s B}\rs{}'\rs{}''\dfrac{\bx'^k(\bx'-\bx'')^i}{|\bs\bx'-\bs\bx''|^3}d^3\bx'd^3\bx''.
\end{align}
The integral in the second line above can be symmetrized to acquire  the structural integral shape shown below,
\begin{equation}
	\Omega^{ik}_{\s B}=-\frac{1}{2}\int_{\s B}\rs{}'\rs{}''\dfrac{(\bx'-\bx'')^k(\bx'-\bx'')^i}{|\bs\bx'-\bs\bx''|^3}d^3\bx'd^3\bx''.
\end{equation}
It is worth noting that the trace of $\Omega^{ij}_{\s B}$ gives the gravitational energy of body $B$. At the end, the second term in \eqref{terms} will read,
\begin{equation}
	\sum_{B\neq A}\dfrac{3\Omega^{ik}_{\s B}}{r^4_{\s AB}}\left(\delta_{ik}n^j_{\s AB}+2\delta_{ij}n^k_{\s AB}-5n^i_{\s AB}n^k_{\s AB}n^j_{\s AB}\right).
\end{equation}

The third term in \eqref{terms} is calculated similar to the first one, but now $\bs x'$ is expanded around $\bs r_{\s B}$ while $\bs x''$ is expanded around $\bs r_{\s C}$. Thus, $\p'_is'\approx - n^i_{\s BC}/r_{\s BC}^2$ and, using the leading part of \eqref{dds} with $\bs x=\bs r_{\s A}$, one gets,
\begin{equation}
\sum_{B\neq A} \sum_{C\neq A,B}	\dfrac{m_{\s B}m_{\s C}}{r_{\s AB}^3r_{\s BC}^2}\left[n_{\s BC}^j- 3(\bs n_{\s AB}\cdot \bs n_{\s BC})n_{\s AB}^j \right].
\end{equation}
By collecting all the previous results,  the external contribution to the gradient of $\Phi^{\s ext}_{\s 7,A} $ can be recast as
\begin{widetext}
	\begin{equation}\label{f19-final}
		\p_j\Phi^{\s ext}_{\s 7,A}=\sum_{B\neq A}\Bigg[\dfrac{2m_{\s A}m_{\s B}}{r^5_{\s AB}}n^j_{\s AB} + \dfrac{3\Omega^{ik}_{\s B}}{r^4_{\s AB}}\left(\delta_{ik}n^j_{\s AB}+2\delta_{ij}n^k_{\s AB}-5n^i_{\s AB}n^k_{\s AB}n^j_{\s AB}\right) + \!\!\! \sum_{C\neq A,B}	\dfrac{m_{\s B}m_{\s C}}{r_{\s AB}^3r_{\s BC}^2}\big(n_{\s BC}^j- 3(\bs n_{\s AB}\cdot \bs n_{\s BC})n_{\s AB}^j \big)\Bigg].\!
	\end{equation}
\end{widetext}

The 4D-EGB correction to the PN-equation of motion of well-separated massive bodies is given by the force expressions \eqref{force}. Substituting in that the results given in \eqref{f20-final} and \eqref{f19-final}, one can explicitly write the acceleration of body $A$ in a system of $N$ bodies as follows,
\begin{widetext}
\begin{align}
	a_A^j =& \ a_{\s A}^j{\rm [GR]} \mp 8\a\sum_{\s B\neq A} \frac{M_{\s B}(M_{\s A}+M_{\s A})}{r_{\s AB}^5} \,n_{\s AB}^j \pm 6\a \sum_{\s B\neq A}\frac{\Omega_{\s B}^{ik}}{r_{\s AB}^4}\left(\delta^{ik}n^j_{\s AB}+2\delta^{ij}n^k_{\s AB}-5n^i_{\s AB}n^k_{\s AB}n^j_{\s AB}\right)\notag\\[1ex]
	& \ \pm 4\a\sum_{\s B\neq A}\sum_{\s C\neq A,B}\frac{M_{\s B}M_{\s C}}{r_{\s AB}^2} \Bigg[\frac{n_{\s AB}^j}{r_{\s AC}^3} - 3\frac{(\bs n_{\s AB}\cdot \bs n_{\s AC})n_{\s AC}^j}{r_{\s AC}^3} - \frac{n_{\s BC}^j}{r_{\s AB}r_{\s BC}^2} + 3\frac{(\bs n_{\s AB}\cdot \bs n_{\s BC})n_{\s AB}^j}{r_{\s AB}r_{\s BC}^2} \Bigg].\label{pn-eom}
\end{align} 
\end{widetext}
In Eq. (\ref{pn-eom}), $a_{\s A}^j{\rm  [GR]}$ represents the standard  GR term to the equation of motion of each body, say body $A$, where each mass term $m$ was replaced by $M$ provided to mass-energy is conserved  [cf. \eqref{mass-energy}], and the corrections introduced by these exchanges are beyond the PN order. In addition to that, the terms proportional to the re-scaled coupling $\hat{\alpha}$ accommodate the new contributions to the acceleration of the $A$ body. In the next section, we will employ  (\ref{pn-eom}) to  perform some astrophysical constraints on the 4D-EGB theory, for instance, by obtaining the orbital periastron shift in a binary system.

\section{Astrophysical constraints}\label{sec:periastron}
\subsection{The periastron advance rate}
To analyze the 4D-EGB contributions to the orbital periastron advance, we consider the two-body problem. We work in terms of relative acceleration,  $\vec{a}=\vec{a}_1 -\vec{a}_2$,  and use the notation $\vec{r}=\vec{r}_1 - \vec{r}_2$, $\hat{n}=\vec{r}/r$, and $m= M_1 +M_2$. We focus on the case of nearly spherical bodies; the approximation $\Omega_{\s N}^{ik}\approx \delta^{ik}\Omega_{\s N}/3$ remains valid. Replacing the latter approximation in \eqref{pn-eom}, one then  obtains a simplified version for the total acceleration,
\begin{equation}\label{acc-rel}
	\vec{a}=\vec{a}{\rm [GR]}\pm \frac{8\a m^2}{r^5}\,\hat{n}.
\end{equation}
The 4D-EGB contribution to the two-body system relative acceleration is of PN order; thus, it can be considered a perturbing acceleration that gives origin to perturbed Keplerian orbits.

Exploiting the method of osculating orbital elements (see Ref. \cite{PoissonWill}), the perturbed motion is interpreted as an instantaneous Keplerian orbit, and the system is always describing an ellipse, but with the orbital elements varying with time. The perturbing acceleration is decomposed in a radial component ${\cal R}$, a component ${\cal W}$ normal to the orbital plane, and a third component ${\cal S}$ transversal to the previous two directions. Equation \eqref{acc-rel} thus shows,
\begin{equation}
    {\cal R}= \frac{8\a m^2}{r^5},\quad {\cal W}={\cal S}=0.
\end{equation}
The variations of the orbital elements can then be directly related to these components.

\subsection{Mercury's periastron advance rate}

Once observations are made in geocentric coordinates, the periastron argument measured is relative to the equinox. Consequently,  it can be recast as $\tilde\omega=\omega+\Omega\cos i$, where $\Omega$ is the angle from the ascending node to the Earth-Sun direction, $\omega$ is the angle between the periastron and the ascending node. At the same time, $i$ indicates the orbital plane inclination angle relative to the ecliptic \cite{Will:1993ns}. Using the fact that, for all planets in the solar system, $i$ is too small, the variation of $\tilde\omega$ then is  determined by the following derivative, 
\begin{equation}\label{dwt}
	\frac{d\tilde{\omega}}{df}= \frac{p^2\sin f}{em(1+e\cos f)^2}\bigg[{\cal S}\,\frac{2+e\cos f}{1+e\cos f} -{\cal R}\cot f\bigg],
\end{equation}
where $f$ is the true anomaly (the angular position of the planet with respect to its periastron direction), $p=a(1-e^2)$ is the semi-latus rectum, whereas $a$ denotes the semi-major axis, and $e$ stands for the eccentricity of the orbit. Within this Keplerian description, the Sun-planet relative position is given by the usual conic equation,
\begin{equation}
    r=\frac{p}{1+e\cos f}.
\end{equation}
Integrating over a complete orbital period gives us the secular change in the periastron position. Using \eqref{acc-rel}, one can calculate the 4D-EGB contribution to this effect, and, after including the GR contribution, one obtains
\begin{equation}\label{delta}
	\Delta\tilde{\omega}=\frac{6\pi m}{p}\left[1\pm \frac{\a}{p^2}\,(4+e^2)\right].
\end{equation}
The result above is precisely found in Ref. \cite{PhysRevD.102.084005}, where the authors considered the planets as described by test particles. Although the general equation of motion \eqref{acc-rel} shows terms depending explicitly on the body's internal structure, through their gravitational energy tensor $\Omega^{ik}$, all those terms cancel each other when assuming nearly spherical bodies. Therefore, it remains to be analyzed in situations where the internal structure of the bodies does contribute to the orbital motion.

The advance per orbit $\Delta \tilde\omega$ can be converted to a rate by dividing it by the orbital period, $P$. The result is given by
\begin{equation}
    \dot{\tilde\omega} = 3\left(\frac{2\pi}{P}\right)^{5/3}\frac{m^{2/3}}{1-e^2} \ \pm \ 3\a\left(\frac{2\pi}{P}\right)^{3}\frac{4+e^2}{(1-e^2)^2}, 
    \label{dot-omega}
\end{equation}
after using Kepler's third law, $P^2=4\pi^2p^3/m(1-e^2)^3$.

Eq. (\ref{dot-omega}) can be used to put some constraints over the parameter $\a$. Data analysis from the MESSENGER mission estimates secular Mercury's periastron precession as $\dot{\tilde\omega}= (42.9799 \pm 0.0009)$\,arcsec per century \cite{Park_2017}. With the help of the orbital parameters $e=0.2056$, $P = 87.97$ days, $M_\odot=1.9891\times 10^{30}$\,kg and $M_{\text{\mercury}}=3.3011\times 10^{23}$\,kg \cite{Stark:2015}, one can find an upper bound limit for the GB coupling, which is given by
\begin{equation}
    \vert\a\vert\lesssim 1.67\times 10^{16}\,\text{m}^2.
\end{equation}
This constraint is two orders of magnitude stronger than the one established in Ref. \cite{PhysRevD.102.084005} due to the use of less precise data on Mercury's orbit.

\subsection{The Double Pulsar periastron shift}

The binary’s orbital motion affects the pulses from a pulsar in a binary system, altering their arrival times at radio telescopes \cite {Taylor:1994zz}. By measuring these arrival times accurately over long periods of time, we can detect slight variations in the orbital motion. In 2003, the double pulsar PSR J0737\,--\,3039A/B was first detected, representing probably one of the best binary systems for studying gravitational effects. Such a claim is  based on the following facts: $i)$ it is the only binary where we can see both pulsars, $ii)$  it is not so far from Earth — nearly $0.61\text{kpc}$, and $iii)$ its orbit is almost “edge-on” to us. All these listed properties indicate why the double pulsar offers a unique opportunity to explore the strong gravity regime and, consequently, any deviation from GR \footnote{A recent study, which can be found in Ref.\cite{AbhishekChowdhuri:2022ora}, discusses the evolution of eccentricity in binary systems within the PN framework, extending beyond Horndeski’s theory. This work also explores its capability to differentiate from GR.}

More recently \citet{Kramer:2021} presented precise determination of periastron precession using 16-year data span of the PSR J0737\,--\,3039A/B.  To obtain stringent bounds using this higher accurate system, it is necessary to derive a new expression for $\dot{\tilde\omega}$, since the inclination $i$ is now close to $90^\circ$. Being $\cos i\approx0$, the variations of $\Omega$ will not affect the periastron advance rate, which, in turn, will lead to a modification in equation \eqref{dwt} proportional to ${\cal W}$. However, due to the radial nature of the 4D-EGB correction to the relative acceleration, the expression \eqref{dot-omega} for $\dot{\tilde\omega}$ remains valid for the double pulsar system. The total mass of the two stars is $2.587M_{\odot}$, their orbit is slightly elliptical with $e=0.0878$, and they complete one revolution in $P=0.10225$ day. The measured precession rate reads $\dot{\tilde\omega}=(16.899323 \pm 0.000013)\,$ degrees per year. Using these data,  it is possible to obtain a stronger constraint on the re-scaled GB coupling, namely,
\begin{equation}
    \vert\a\vert\lesssim 1.47\times 10^{11}\,\text{m}^2.
\end{equation}
By comparing the bounds found in Sections VIB and VIC, it is fair to state that the double pulsar system imposes a much stronger constraint by reducing the bounds on $\hat{\alpha}$ by five orders of magnitude.  Nevertheless, the above estimation on GB coupling represents an interesting improvements in regard to the previous  best estimates, $|\hat{\alpha}|<10^{15}\, \text{m}^2$ \cite{PhysRevD.102.084005}.

\subsection{Other observational bounds}

Let's discuss the improvement introduced on the $\hat{\alpha}$ coupling's bound concerning other observational tests reported in the literature in different regimes and sectors of the 4D-EGB theory. 
Even though the constraints coming from the double pulsar periastron shift enhanced the bounds on $\hat{\alpha}$ by several orders of magnitude, it fell short regarding the bounds emerging in the strong-field regime provided by the binary black hole system \cite{PhysRevD.102.084005}. In other words,  $\hat{\alpha}_{BBH}<\hat{\alpha}_{DPP}<10^{10}\,{\rm{m}}^{2}$. The latter situation is considerably improved once is assumed that lightest component of the GW190814 event could potentially correspond to  stellar black hole  with mass $M= 2.59^{+0.08}_{-0.09}M_{\odot}$, yielding $\hat{\alpha}_{GW190814}< 59 \times 10^{6}\,{\rm{m}}^{2}$ \cite{Charmousis:2021npl}. When we account for an extra-dark radiation term at both the background and perturbative levels in cosmology, we can constrain it using the ACTPol data alone.  This gives us $\hat{\alpha}C^{2}= (-9\pm 6)\times 10^{-6}H^{2}_{0}$, where $\hat{\alpha}C^{2}=\Omega^{\rm{extra}}_{\rm{rad}}H^{2}_{0}$ \cite{Zanoletti:2023ori}. If we ignore the geometrical dark radiation term, however, we get a less robust estimate on $\hat{\alpha}<   10^{36}\,{\rm{m}}^{2}$  \cite{Fernandes:2022zrq}. 
Given that nucleosynthesis started at an energy scale of $E \sim \text{Mev}$ when the universe is dominated by radiation,  we can estimate that $\hat{\alpha}< 10^{18}\,{\rm{m}}^{2}$  \cite{Fernandes:2022zrq}. However, this bound is much weaker than the double pulsar’s best estimates.

\section{The \texorpdfstring{$\boldsymbol{\beta}$}{b} parameter and the new EGB parameter: an extended PPN version}\label{sec:beta}

The previous sections showed that 4D-EGB corrections to GR post-Newtonian metric, does not change how light travels or how energy and momentum are conserved in the weak gravity limit up to the first PN order. Within the PPN formalism, these results demonstrate that the physical meaning associated with parameters  $\gamma$, $\zeta$'s, and $\alpha$'s are not influenced by the new potential $\PhiG$. Consequently, once it was shown that photon geodesics and total energy and momentum expressions are the same as in GR, it was possible to conclude that $\gamma=1$ and all $\zeta$'s and $\alpha$'s are null. The remaining PPN parameters to be determined are $\beta$ and $\xi$. To move forward with the determination of the PPN parameters, one assumes the knowledge of the already determined parameters in juxtaposition with the general PPN expression for $\Delta\omega$, namely,
\begin{align}
    \Delta\tilde\omega_{\rm PPN} = \frac{2\pi m}{p}(4 -\beta),\label{dw-ppn}
\end{align}
one might be tempting, by comparison with \eqref{delta},  to define an effective $\beta$ parameter that would encompass the 4D-EGB corrections to GR. However, we understand that such treatment is not appropriate once it would lead to a system-dependent parameter, a rather strange feature for the PPN formalism. More drastically, if an effective parameter is defined as a system-dependent constant, one could not simply substitute the original parameter with its effective version in each observable physical phenomenon. As an example,  we emphasize that if an effective $\beta$ were defined through \eqref{dw-ppn} and \eqref{delta}, this new parameter would not have any influence on the so-called Nordtvedt effect: the violation of the weak equivalence principle due to explicit contributions of self-gravitational energy to a body's inertial and gravitational mass.

The Nordtvedt effect can be tested for the Earth-Moon system by studying its motion in the Sun's gravitational field. If there is any difference between the Earth's free-fall acceleration and the Moon's towards the Sun, this effect will be parametrized by the Nordtvedt parameter,
\begin{equation}
\eta= 4\beta -4 -\frac{10}{3}\,\xi,\label{nordtvedt}
\end{equation}
where $\xi$ is the PPN parameter related to the existence of preferred-locations effects, and, once again, we use $\gamma=1$ and $\zeta_i=\alpha_i=0$.
It has been shown that $\PhiG$ does not bring any dependence in the acceleration of a body $A$ with its internal structure up to the Newtonian order [cf. Eq. \eqref{a}]. Consequently,$\eta=0$ and no Nordtvedt effect is present, just like in GR. Therefore, a simple substitution of $\beta$, in \eqref{nordtvedt}, by an effective parameter obtained from the periastron advance would lead to erroneous conclusions for any theory.

The correct approach to tackle the issue above would be to propose a new parameter, say $\tilde{\beta}$, to quantify the dynamical effects associated with $\PhiG$. In this extended PPN version (EPPN), the Nordtvedt parameter will remain the same, while the periastron advance per orbit will be given by
\begin{align}
    \Delta\tilde\omega_{\rm EPPN}=\Delta\tilde\omega_{\rm PPN}+\frac{\tilde\beta}{p^2}\,(4+e^2).
\end{align}
Within general relativity, for instance, $\beta=1$ and $\tilde\beta=0$, while in the case of 4D-EGB theory one has $\beta=1$ and $\tilde\beta=\pm\,\a$. Hence, with $\eta=0$ one would also obtain $\xi=0$. However, it is important to emphasize that, within this extended PPN approach, $\beta$ alone does not determine the post-Newtonian periastron advance effect. For instance, the $\beta=1$ result is insufficient to ensure agreement with Mercury's periastron advance test. Still, it is essential to guarantee no violations of the weak equivalence principle in the case of fully conservative theories.

\section{Summary}\label{sec:conclusion}

We presented a comprehensive analysis of the post-Newtonian version of the regularized four-dimensional Einstein-Gauss-Bonnet gravitational theory. Our investigation involved expanding the metric and the extra scalar field around the Minkowski background and the constant scalar field value, respectively. To model the matter content, we adopted a perfect fluid approach. The post-Newtonian metric components are detailed in equations (\ref{g002}) through (\ref{gij2}). Notably, the primary correction emerges at fourth order in the $g_{00}$ component. Unlike previous works within the post-Newtonian formalism, we utilized the harmonic gauge  and expressed the gravitational potential in terms of the conserved density.

Starting from conservation laws, it was shown that there is not violation of energy and/or momentum conservation within the PN limit, confirming that the four of  $\zeta_{i}$ PPN parameters  and $\alpha_{3}$ vanish. Moreover, the expressions for energy and momentum showed that parameters $\alpha_1$ and $\alpha_2$ are also zero; this highlights that the regularized four-dimensional Einstein-Gauss-Bonnet gravitational theory lacks any preferred frame effects.

We investigated the equation of motion for massive bodies and derived the general expression for a body within a system of N bodies. Furthermore, we recovered the standard GR term and identified several contributions arising from the Gauss-Bonnet (GB) term. The latter discovery involves lengthy computations, but it was essential for imposing  physical bounds on the re-scaled GB coupling.

From an observational perspective, we examined the two-body problem for nearly spherical objects and utilized the method of osculating orbital elements to derive the PN corrections in the periastron advance rate due to the Gauss-Bonnet  term. By analyzing data from the {MESSENGER} mission applied to secular Mercury’s periastron precession, we arrived at an updated estimate for the GB coupling:  $\vert\a\vert\lesssim 1.67\times 10^{16}\,\text{m}^2$; consistent with previous results \cite{PhysRevD.102.084005}.  Further,  we  focused on the double pulsar {J0737\,--\,3039A/B},  comprising two active radio pulsars  in a binary system with a period of 2.45 hours and a mild eccentricity of $e=0.088$. Taking into account the data collected on extensive pulse timing experiments over 16 years on six different radio telescopes around the world, this experiment reported the precession rate for the double pulsar system, $\dot{\tilde\omega}=(16.899323 \pm 0.000013)\,$ degrees per year (see Ref. \cite{Kramer:2021} for further details). We proceeded by using this precession rate of the double pulsar system to put a tighter constraint on the GB coupling, yielding $\vert\a\vert\lesssim 1.47\times 10^{11}\,\text{m}^2$; reducing the former bound in five orders of magnitude. We also compared other astrophysical and cosmological bounds on $\hat{\alpha}$ to contextualize the newly updated results reported here. 

Finally, we discussed, to a certain degree, the $\beta$ PPN parameter identification and the introduction of a new EGB parameter.  This analysis focused on the different consequences of misreading the $\beta$ PPN parameter within a modified gravity scenario such as the 4D-EGB model. 

\section{Acknowledgments}
M.G.R. was partially supported by FAPES (Brazil) and CNPq (Brazil). The authors acknowledge interesting comments from Prof. K. Hinterbichler and Prof. A. Bhattacharyya.

\appendix

\section{The PPN formalism -- a synthesis}\label{appendix}

In order to distinguish between alternative theories of gravity within the post-Newtonian regime, the PPN formalism assumes a generic metric expansion under reasonable (but restrictive) assumptions over the possible gravitational potentials to be present.

The original PPN metric, as presented in Ref. \cite{Will:1993ns}, is written in the so-called standard gauge. Nowadays, it is more usual for PN analysis to work with the harmonic gauge. Although the gauge choice is a convenience matter, the harmonic gauge usage allows to consistently connect the post-Newtonian approximation to eh wider scopus of a post-Minkowskian theory (see Ref. \cite{PoissonWill} for details). In the harmonic gauge, the PPN metric reads,

\begin{align}
	g_{00} = -&1 + 2U - 2\beta U^2 + (2 \gamma +2+\alpha_3 +\zeta _1-2 \xi ) \Phi_1 \,+\notag\\[1ex]
	+& \, 2(3 \gamma -2\beta+1+\zeta _2+ \xi ) \Phi_2 \ + 2(1+\zeta _3 ) \Phi_3 \,+ \notag\\[1ex]
      +& \, 2(3 \gamma +3\zeta _4-2 \xi ) \Phi_4 - (\zeta _1-2 \xi )\Phi_6 \, - \notag\\[1ex]
      -& \, 2\xi \Phi_W + (1+\alpha_2 -\zeta_1+2\xi)\p_{tt}\chi \nonumber \\[2ex]
	g_{0i} = -&(2\gamma +2+\alpha_1/2) V_i \label{34} \\[2ex]
	g_{ij} = \ (&1+2\gamma \,U)\, \delta_{ij}\,.\nonumber
\end{align}
\begin{table}[h!]
  \caption{The PPN Parameters physical meaning.}
  \label{tab:parameters}
  \renewcommand{\arraystretch}{1.5}
  \centering
  \begin{tabular}{p{1.5 cm}|p{4.2 cm}|c}
    \hline \hline
    Parameter & Physical meaning associated & Value in GR \\
    \hline
    $\gamma$ & light motion tests & 1\\
    \hline
    $\beta$ & Periastron shift in a binary system & 1\\
    \hline
    $\xi$ & Preferred-location effects & 0\\
    \hline
    $\alpha_1$, $\alpha_2$ & Preferred-frame effects & all null\\
    and $\alpha_3$ & \\
    \hline
    $\zeta_1$, $\zeta_2$, $\zeta_3$, & Violation of conservation & all null \\
    $\zeta_4$ and $\alpha_3$ & of total momentum & \\
    \hline \hline
  \end{tabular}
  \renewcommand{\arraystretch}{1.0}
\end{table}

The organization of the PPN parameters in the metric above is done in such a way that is possible to give to each one of them a specific physical meaning which associates them to a property or a measurable effect. Table \ref{tab:parameters} summarize the significance of each PPN parameter and the current bounds on PPN parameters are shown in Table \ref{tab:bounds}.

\begin{table}[h!]
  \caption[Current limits on the PPN parameters.]{Current limits on PPN parameters (from Ref. \cite{Will:2014kxa}). }
  \label{tab:bounds}
  \renewcommand{\arraystretch}{1.5}
  \centering
  \begin{tabular}{c|r||c|c}
    \hline \hline
    Parameter & Limit\phantom{AA} & 
    Parameter & Limit \\
    \hline
    $\gamma-1$ & \ ${\scriptstyle \lesssim\,} 2.3 \times 10^{-5 \phantom{0}}$ & $\alpha_3$ &   \ ${\scriptstyle \lesssim\,}4 \times 10^{-20}$ \\
    $\beta-1$ & \ ${\scriptstyle \lesssim\,}8 \times 10^{-5 \phantom{0}}$ &$\zeta_1$ &  \ ${\scriptstyle \lesssim\,}2 \times 10^{-2 \phantom{0}}$ \\
    $\xi$ & \ ${\scriptstyle \lesssim\,}4 \times  10^{-9 \phantom{0}}$ & $\zeta_2$ & \ ${\scriptstyle \lesssim\,}4 \times 10^{-5 \phantom{0}}$ \\
    $\alpha_1$ & \ ${\scriptstyle \lesssim\,}7 \times 10^{-5 \phantom{0}}$ & $\zeta_3$ & \ ${\scriptstyle \lesssim\,}10^{-8 \phantom{0}}$ \\
    $\alpha_2$ & \ ${\scriptstyle \lesssim\,}2 \times 10^{-9 \phantom{0}}$ & $\zeta_4$ &  \multicolumn{1}{c}{---} \\
    \hline \hline
  \end{tabular}
\end{table}

When a theory does not present violations of energy and momentum at the PN regime, all $\zeta$'s and $\alpha_3$ vanish. The expressions for the conserved energy and momentum are then reduced to
\begin{equation}
E\equiv\int\left(\frac{1}{2}\rs v^{2}  + \rs\Pi - \frac{1}{2}\rs U \right)d^{3}x.
\end{equation}
\begin{align}
    P_j=& \int\rho^{\ast}v_{j}\left(  1+\frac{1}{2}v^{2}-\frac{1}{2}U+\Pi+\frac
{p}{\rho^{\ast}}\right)  d^{3}x\label{Momentum-PPN}\\
&-\frac{1}{2}\int\rho^{\ast}\left[(1+\alpha_2)\Phi_{j}+\frac{1}{2}(\alpha_1-\alpha_2)V_{j}\right]d^{3}x,\nonumber
\end{align}

\bibliography{AllMyRefs}{} 

\end{document}